%% file: main.tex
\PassOptionsToPackage{table,dvipsnames}{xcolor}
\documentclass[sigconf]{acmart}
\usepackage[table]{xcolor}
\usepackage{tikz}

\usepackage{algorithm}
\usepackage{algorithmic}
\definecolor{darkpurple}{RGB}{102, 51, 153}
\usepackage{enumitem}
\usepackage{threeparttable}
\usepackage{multirow}
\usepackage{array}
\usepackage{amsmath}
\usepackage{pifont}

\usepackage{balance}

\usepackage{amssymb}

\definecolor{cmarkcolor}{HTML}{58a65d}

\newcommand{\cmark}{\textcolor{cmarkcolor}{\ding{51}} }  
\newcommand{\warn}{\textcolor{orange}{\ding{115}} } 

\usepackage[edges]{forest}
\definecolor{hidden-draw}{RGB}{0,0,0}
\definecolor{hidden-pink}{rgb}{0.98, 0.94, 0.75}
\definecolor{level0}{rgb}{0.67, 0.88, 0.69}
\definecolor{level1}{rgb}{0.98, 0.92, 0.84}
\definecolor{level2}{rgb}{0.8, 0.8, 1.0}
\definecolor{level3}{rgb}{1.0, 0.71, 0.76}
\definecolor{level4}{rgb}{0.49, 0.99, 0.0}

\newlength\savedwidth

\newlength\savewidth
\newcommand\shline{\noalign{\global\savewidth\arrayrulewidth
                            \global\arrayrulewidth 1.5pt}%
                   \hline
                   \noalign{\global\arrayrulewidth\savewidth}}

\definecolor{mygray}{gray}{0.5}
\definecolor{background_gray}{gray}{0.9}
\definecolor{lawngreen}{rgb}{0.49, 0.99, 0.0}
\definecolor{pink}{rgb}{1, 0, 0.5}
\definecolor{airforce}{rgb}{0.36, 0.54, 0.66}

\hypersetup{
    linkbordercolor=airforce,
    urlbordercolor=pink,
    citebordercolor=green,
}
\AtBeginDocument{%
  \providecommand\BibTeX{{%
    \normalfont B\kern-0.5em{\scshape i\kern-0.25em b}\kern-0.8em\TeX}}}

\begin{document}


\title[Foundation Models for Spatio-Temporal Data Science: A Tutorial and Survey]{Foundation Models for Spatio-Temporal Data Science: \\ A Tutorial and Survey}


\author{Yuxuan Liang\textsuperscript{\rm 1}, Haomin Wen\textsuperscript{\rm 2,1}, Yutong Xia\textsuperscript{\rm 3}, Ming Jin\textsuperscript{\rm 4},  Bin Yang\textsuperscript{\rm 5}, \\ Flora Salim\textsuperscript{\rm 6}, Qingsong Wen\textsuperscript{\rm 7}, Shirui Pan\textsuperscript{\rm 4}, Gao Cong\textsuperscript{\rm 8}} 
\affiliation{%
  \institution{
  \textsuperscript{\rm 1}The Hong Kong University of Science and Technology (Guangzhou) \hspace{0.1em}
  \textsuperscript{\rm 2}Carnegie Mellon University  \\ 
  \textsuperscript{\rm 3}National University of Singapore \hspace{0.1em} 
  \textsuperscript{\rm 4}Griffith University \hspace{0.1em}
  \textsuperscript{\rm 5}East China Normal University \\ 
  \textsuperscript{\rm 6}University of New South Wales \hspace{0.1em} 
  \textsuperscript{\rm 7}Squirrel Ai Learning, USA \hspace{0.1em} 
  \textsuperscript{\rm 8}Nanyang Technology University 
  }
  \city{} 
  \state{}
  \country{}
}

\email{{yuxliang,yutong.x}@outlook.com, {wenhaomin.whm,mingjinedu,qingsongedu}@gmail.com}
\email{flora.salim@unsw.edu.au, byang@dase.ecnu.edu.cn, s.pan@griffith.edu.au, gaocong@ntu.edu.sg}

\renewcommand{\shortauthors}{Yuxuan Liang et al.}

\begin{abstract}
Spatio-Temporal (ST) data science, which includes sensing, managing, and mining large-scale data across space and time, is fundamental to understanding complex systems in domains such as urban computing, climate science, and intelligent transportation. Traditional deep learning approaches have significantly advanced this field, particularly in the stage of ST data mining. However, these models remain task-specific and often require extensive labeled data. Inspired by the success of Foundation Models (FM), especially large language models, researchers have begun exploring the concept of Spatio-Temporal Foundation Models (STFMs) to enhance adaptability and generalization across diverse ST tasks. Unlike prior architectures, STFMs empower the entire workflow of ST data science, ranging from data sensing, management, to mining, thereby offering a more holistic and scalable approach. Despite rapid progress, a systematic study of STFMs for ST data science remains lacking. This survey aims to provide a comprehensive review of STFMs, categorizing existing methodologies and identifying key research directions to advance ST general intelligence.
\end{abstract}



\maketitle

%
\vspace{-1em}
\section{Introduction} 

Humans live in a world shaped by the dynamic interplay of countless elements across space and time. \textit{Spatio-Temporal (ST) Data}, which refer to data that encapsulate ST phenomena, track the evolution of objects or events across locations and time \cite{atluri2018spatio}, such as meteorological records, traffic patterns, and human traces. These data are frequently sourced from a wide array of platforms, ranging from IoT devices, GPS sensors, social media, to remote sensing. 

Within this context, \textbf{Spatio-Temporal Data Science} focuses on \textit{sensing}, \textit{managing}, and \textit{mining} these datasets to uncover patterns, understand complex systems, and predict future dynamics. Motivated by its transformative potential, this field addresses critical challenges across urban environments and even the entire planet, enabling decision-making and fostering innovations that contribute to building smarter, more sustainable, and resilient systems \cite{zheng2014urban}.

In the era of deep learning, the community has primarily concentrated on \emph{spatio-temporal representation learning}, as a fundamental step of ST data mining \cite{wang2020deep}. Key advancements include the development of Spatio-Temporal Graph Neural Networks (STGNN) \cite{jin2023spatio} and transformer-based architectures, which have shown remarkable success in tasks such as traffic forecasting \cite{xu2020spatial,liang2022mixed}, air quality prediction \cite{liang2023airformer}, and human mobility analytics \cite{wang2024spatiotemporal}. STGNNs integrate Graph Neural Networks (GNN) with temporal learning modules (e.g., GRU \cite{li2017diffusion,bai2020adaptive}, TCN \cite{wu2019graph,wu2020connecting}) to model ST correlations, while transformer models leverage self-attention mechanisms \cite{liang2018geoman,guo2019attention,zheng2020gman} to process complex dependencies across space and time. Additionally, there has been significant research on \emph{self-supervised learning} \cite{liu2022contrastive,ji2022self,li2024gpt}, where models are trained to extract powerful representations with minimal reliance on large annotated datasets. 

\begin{figure}[!t]
    \centering
    \includegraphics[width=0.98\linewidth]{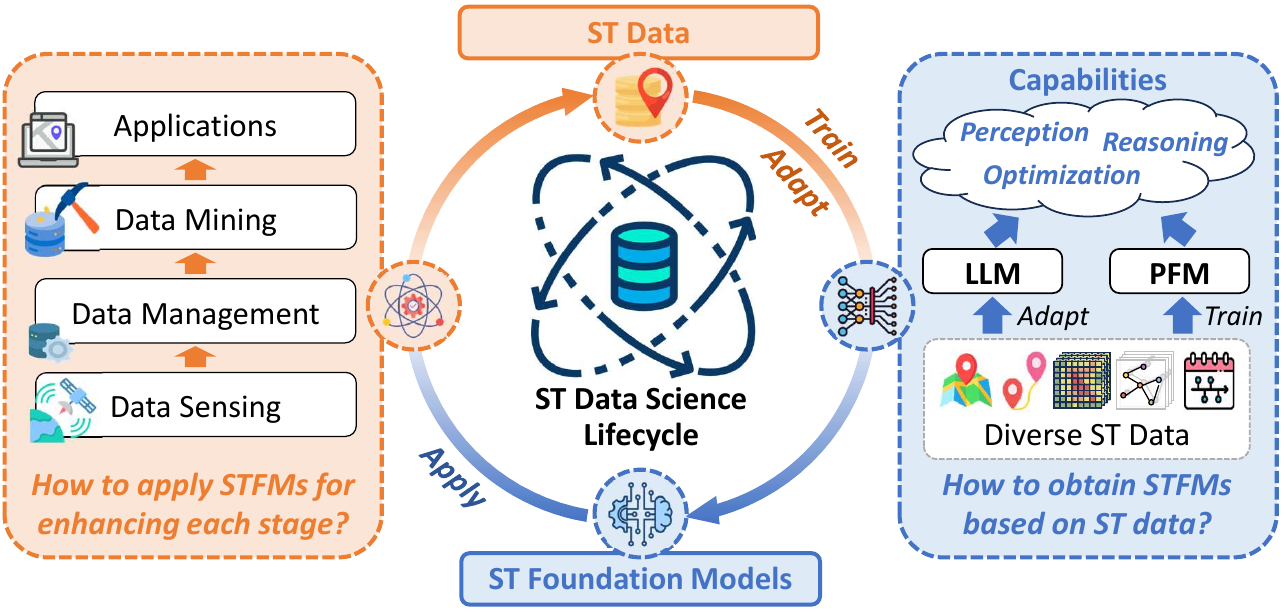}
    \vspace{-0.5em}
    \caption{ST Foundation Models (STFM), which include LLM and PFM, are pretrained with or applied to diverse ST data, with the abilities of perception, optimization, and reasoning. STFMs can, in turn, enhance each stage of ST data science.}
    \label{fig:intro}
    \vspace{-1.5em}
\end{figure}
Driven by the success of Foundation Models (FM), especially Large Language Models (LLM), researchers have recently begun exploring the concept of \textbf{Spatio-Temporal Foundation Models} (STFM) \cite{liang2024foundation,zhang2024towards,goodge2025spatio}. By harnessing LLMs, it becomes possible to develop more generalized, adaptable solutions that can be fine-tuned for specific tasks with minimal data. Another prominent approach involves pretraining FMs (denoted as PFM) on cross-domain ST data and adapting them for particular domains. In contrast to previous architectures (e.g., STGNNs), STFMs integrates the capabilities of \emph{perception}, \emph{reasoning} and \emph{optimization}, which not only promises to revolutionize ST data mining, \emph{but also empowers other stages of ST data science}, such as ST data sensing and management (See Figure \ref{fig:intro}). This shift has the potential to enhance the scalability and efficiency of ST applications, offering a more holistic approach to addressing challenges in urban computing, climate science, etc.

Despite their rapid advancements, a systematic analysis of STFMs across the \textbf{entire workflow} of ST data science remains lacking. First, prior surveys have primarily focused on utilizing LLMs as the key tool for ST data mining  \cite{liang2024foundation,jin2023large,zhang2024towards,goodge2025spatio}, leaving a significant gap in understanding how these models can be integrated throughout the entire process, i.e., \emph{with less focus placed on their role in the earlier stages of sensing and management}. Second, they predominantly examine the applications of STFMs to numerical problems (e.g., forecasting, imputation) while \emph{overlooking their role in inferential problem-solving} such as decision-making systems.

To bridge these gaps, this paper aims to provide a more comprehensive survey of STFMs across all stages of ST data science, spanning data sensing, management, and mining (see Figure \ref{fig:intro}). For example, LLMs can enhance ST data sensing by actively processing citizen reports, optimizing participatory sensing strategies, and generating synthetic data at scale. In terms of data management, they can automate data cleaning tasks, construct meaningful knowledge graphs for data integration, and facilitate more efficient retrieval of cross-modal datasets. Beyond these stages, our survey also explores how STFMs support a broader range of downstream applications, including numerical and inferential problems. Through this endeavor, we seek to illuminate an overall vision of STFMs, thereby enhancing comprehension regarding their potential to optimize ST data science, fostering more integrated and adaptable solutions.

\begin{table}[!t]
\centering
\setlength\tabcolsep{5 pt}
\caption{Our survey vs. related surveys on FMs for learning ST data, such as locations (L), trajectories (T), events (E), ST rasters (R), and ST graphs (G). The applications (App.) include numerical (N) and inferential (I) problems.}
\vspace{-1em}
\resizebox{1\linewidth}{!}{
\begin{tabular}{c|c|c|cccc|c}
\shline
\textbf{Survey} & \textbf{Year} & \textbf{Venue} & \textbf{Sensing} & \textbf{Manage.} & \textbf{Mining} & \textbf{App.} & \textbf{Data} \\
             \hline
Jin et al. \cite{jin2023large}  &   2023 & - &      &            & \ding{52}  & N  &   R,G  \\
Jiang et al.\cite{jiang2024empowering}  &   2024 & IJCAI &      &            & \ding{52}  & N    &  R,G    \\
Liang et al. \cite{liang2024foundation} &  2024 &  KDD &      &            & \ding{52}  & N    &   T,E,R,G    \\
Zhang et al. \cite{zhang2024towards} &    2024 &  KDD &    &            & \ding{52}  & N,I  &    L,T,E,R,G         \\
Goodge et al. \cite{goodge2025spatio}  &  2025 &  - &      &            & \ding{52}  & N    & T,E,R,G  \\
\hline
\hline
\rowcolor{background_gray}  
\textbf{Ours}         & 2025 & - &\ding{52}        & \ding{52}          & \ding{52}  & N,I   &   L,T,E,R,G  \\
\shline
\end{tabular}
}\label{tab:comparison}\vspace{-1.5em}
\end{table}

Meanwhile, we systematically investigate the key methodologies of STFMs for modeling a variety of ST data. We begin by categorizing existing STFMs into two main classes: \emph{LLMs} and \emph{Pretrained Foundation Models (PFMs)}. For LLMs, which are pretrained on linguistic data, we focus on their usage as a zero-shot \cite{gruver2023large} or few-shot learner \cite{jin2023timellm,li2024urbangpt}, where various prompting and fine-tuning strategies have been explored, respectively. For PFMs, which are trained from scratch based on cross-domain ST data \cite{yuan2024unist,zhu2024unitraj,hao2024urbanvlp}, we examine their neural architectures, pretraining methods, and their adaptability to different types of ST data, including location data, trajectory data, events, ST raster data, and ST graph data. 

In summary, our major contributions lie in three aspects:
\vspace{-0.3em}
\begin{itemize}[leftmargin=*]
    \item \textbf{Comprehensive and up-to-date survey}: We provide the first comprehensive and modern survey of FMs across the entire workflow of ST data science, covering data sensing, management, and mining. We also explore a broader range of downstream tasks and data types compared to most existing surveys (See Table \ref{tab:comparison}).
    
    \item \textbf{Vision and Methodologies}: We propose a vision for STFMs, identifying key capabilities essential for their success, and discuss current methodologies for implementing these abilities in detail.
    
    \item \textbf{Future directions}: We highlight promising directions for advancing ST data science with foundation models, encouraging further research and exploration in this emerging field.
\end{itemize}

\newpage
\noindent \textbf{Paper Organization.} The remainder of this paper is organized as follows: Sec. 2 provides essential background on FMs and ST data. Sec. 3 and 4 present a taxonomy of STFMs regarding the workflow and methodologies, respectively. Sec. 5 offers concluding remarks, and Appendix \ref{sec:future} highlights promising avenues for future research.

\vspace{-0.5em}
\section{Background} 
\noindent \textbf{Foundation models.} FMs are deep neural networks trained on vast datasets, enabling them to acquire broad, cross-domain knowledge and exceptional adaptability~\cite{hurst2024gpt}. Unlike earlier task-specific models, FMs can be efficiently fine-tuned with relatively small amounts of task-specific data, offering remarkable flexibility, effectiveness, and cost efficiency. Pioneering attempts like BERT \cite{kenton2019bert} and GPT-3 \cite{brown2020language} have reshaped natural language processing. More recent models, e.g., GPT-4o~\cite{hurst2024gpt} and DeepSeek-R1~\cite{guo2025deepseek}, further push the frontiers of generative capabilities, enabling more nuanced reasoning, robust domain adaptation, and improved context-awareness in diverse tasks. In ST domains, recent FMs like Time-MoE~\cite{shi2024time}, Chronos~\cite{ansari2024chronos}, and UniST~\cite{yuan2024unist} have made remarkable strides in time series analysis and universal ST forecasting, while UniTraj \cite{zhu2024unitraj} serves as a versatile foundation for various trajectory-related tasks. Inspired by these successes, this survey delves into the utilization of FMs in the entire workflow of ST data science, covering data sensing, management, and mining.

\begin{figure}[!b]
    \vspace{-1.5em}
    \centering
    \includegraphics[width=0.92\linewidth]{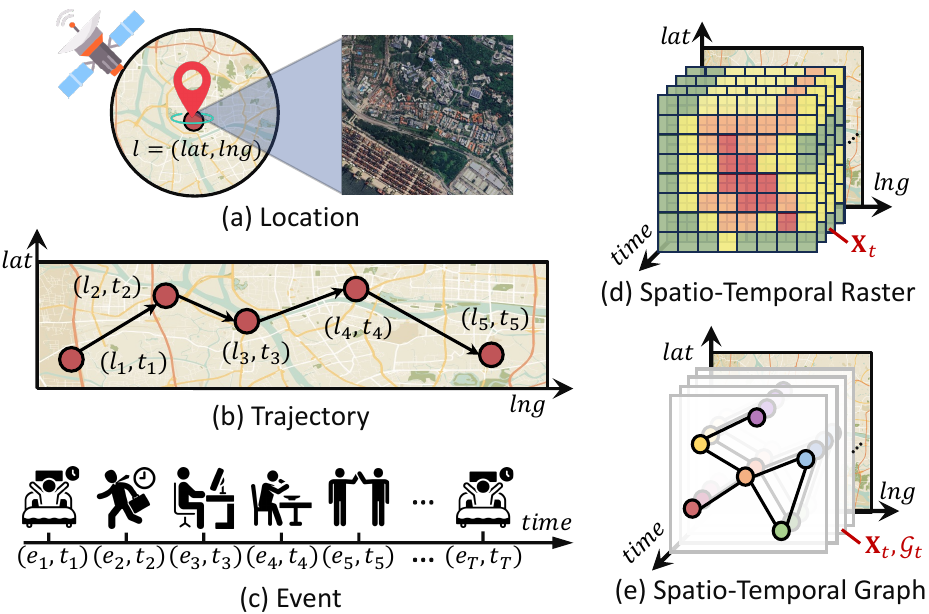}
    \vspace{-1em}
    \caption{Illustration of various types of ST data.}
    \label{fig:data_type}
\end{figure}

\vspace{0.3em}
\noindent \textbf{Formulation of Spatio-Temporal Data.} \label{sec:STD_data}
ST data refer to datasets that integrate spatial (location-based) and temporal (time-based) information, capturing dynamic patterns and relationships over space and time. Figure~\ref{fig:data_type} depicts the basic ST data structures discussed in this survey, including locations, trajectories, events, ST rasters, and ST graphs. Their definitions are delineated as follows.

\textit{Definition 1 (Location).} 
A \textit{location} refers to a fixed spatial point or object in a geographical space, represented by the geospatial coordinates $l \in \mathbb{R}^2$, i.e., latitude and longitude. It is often profiled by the corresponding satellite image, street-view image, and descriptions.


\textit{Definition 2 (Trajectory).} A \textit{trajectory} is a sequence of time-ordered locations that describe the movements of an object in the geographical space. It can be formulated as 
$\mathcal{T} = p_1 \rightarrow p_2 \rightarrow \cdots \rightarrow p_T$,
where $p_i = (l_i, t_i) $, and $l_i$ denotes the object's location at time $t_i$.

\textit{Definition 3 (Event).}  An \textit{event} sequence is a series of timestamped events, denoted as 
$\mathcal{E} = v_1 \rightarrow v_2 \rightarrow \cdots \rightarrow v_T$, describing the progress of actions or occurrences, where $ v_i = (e_i, t_i)$ and $e_i \in \mathbb{R}^d $ is an event and $t_i$ denotes the time when $e_i$ occurs.

\textit{Definition 4 (Spatio-Temporal Raster).}  
An \textit{ST raster} can be denoted as $\mathcal{X} = < \mathbf{X}_1, \mathbf{X}_2, \dots, \mathbf{X}_T > \in \mathbb{R}^{H \times W \times T \times D}$, where $\mathbf{X}_t \in \mathbb{R}^{H \times W \times D}$ denotes the signals collected from $N=HW$ evenly distributed locations at time $t$, each characterized by $D$ feature attributes. 

\textit{Definition 5 (Spatio-Temporal Graph).}  
An \textit{ST graph} extends the ST raster to be $\mathcal{X} = <\mathbf{X}_1, \mathbf{X}_2, \dots, \mathbf{X}_T> \in \mathbb{R}^{N \times T \times D}$ by explicitly incorporating spatial correlations with a graph $\mathcal{G}_t = (V, E_t, \mathbf{A}_t)$ when $N$ locations are not uniformly distributed. Here $V$ is the set of nodes, $E_t$ is the set of edges, and $\mathbf{A}_t \in \mathbb{R}^{N \times N}$ is the adjacency matrix at time $t$. The size of $V$ is usually static.


\begin{figure}[!b]
    \centering
     \vspace{-1em}
    \includegraphics[width=0.99 \linewidth]{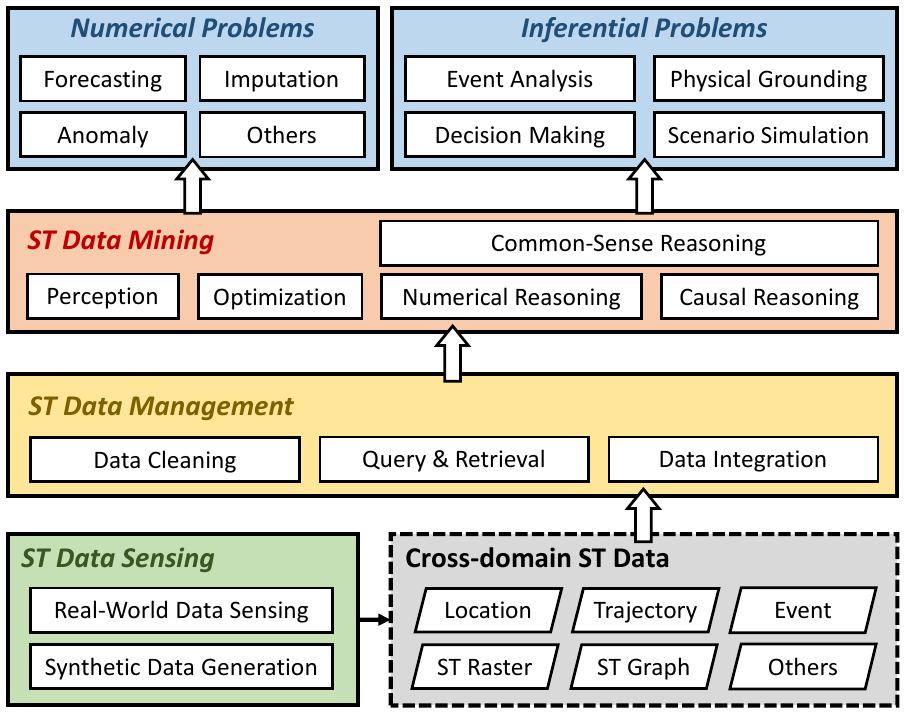}
    \vspace{-1em}
    \caption{The framework of STFMs for ST data science.}
    \label{fig:workflow}
\end{figure}
\section{The Workflow Perspective} 
As shown in Figure \ref{fig:workflow}, we examine STFMs from a holistic, bottom-up perspective, emphasizing their composition across four key aspects: 
\begin{itemize}[leftmargin=*]
    \item \textbf{ST Data Sensing} refers to the acquisition of data that varies over both space and time from diverse resources (e.g., sensors, satellites, social media), to capture dynamic environmental, geographic, or social phenomena. We also consider synthetic data generation for enhancing data diversity and quantity.

    \item \textbf{ST Data Management} focuses on storing, indexing, and organizing these large-scale, heterogeneous ST datasets, incorporating strategies like distributed architectures for efficient retrieval and integration. FMs can enhance this process by facilitating data cleaning, query \& retrieval, and data integration.

    \item \textbf{ST Data Mining} involves learning and analyzing ST data that varies across both space and time to uncover patterns, trends, and relationships, using data mining (DM), deep learning (DL) techniques, or the newly-proposed STFMs with strong capabilities in perception, optimization, and reasoning.

    \item \textbf{Downstream Applications}: This stage harnesses the above insights from ST data to drive real-world applications, ranging from numerical problems to inferential problems, where informed actions and policies are formulated. 
\end{itemize}
By examining these four aspects, we can better understand how STFMs advance from raw data acquisition to high-level service providing, ultimately enabling more intelligent, adaptable, and impactful solutions. We will detail each stage in the following sections.



\subsection{Spatio-Temporal Data Sensing} 

FMs revolutionize ST data sensing from two complementary aspects: \emph{real-world data sensing}, which involves collecting data from physical sources, and \emph{synthetic data generation}, which creates synthetic ST data through foundation models.

\vspace{-0.3em}
\subsubsection{Real-World Data Sensing}
Advances in sensing and data acquisition technologies have led to the generation of vast amounts of ST data. FMs are increasingly applied in human-centric active sensing, particularly in the context of citizen reporting for urban and environmental monitoring \cite{hou2025urban}. These models act as powerful agents for collecting and processing real-time data from citizens, enabling the efficient handling of ST data \cite{dos2024identifying,chhikara2024lamsum,mittelstadt2024large}. For example, citizens might constantly report incidents, environmental changes, or social events through text or voice \cite{zheng2014urban}. By understanding these reports, LLMs can categorize, prioritize, and trigger appropriate responses for various urban issues, from traffic congestion to environmental hazards. This enhances the decision-making process by continuously updating their models with new data streams. Thus, LLMs are not just passive analytical tools but active participants that help make urban environments more responsive and adaptive to citizen inputs, transforming traditional citizen feedback into actionable knowledge, enabling more sustainable and resilient cities.

FMs can also function as intelligent schedulers or simulate multi-agent systems to optimize the recruitment and coordination of participants for crowdsensing, particularly under budget constraints \cite{hou2025urban,zhu2024conversational,wu2024autonomous}. By analyzing ST data and understanding context, LLMs can identify regions and times where crowdsensing efforts will yield the most valuable information. They dynamically recruit participants based on proximity, availability, and past contributions, reducing redundant data collection. Additionally, LLMs simulate multiple agents interacting in real time, ensuring the efficient distribution of sensing tasks across a network of citizens or devices \cite{zhu2024prototype}. This strategic scheduling and agent-based coordination maximize coverage while minimizing costs, ensuring that crowdsensing delivers valuable, real-time insights under budgetary constraints.

\vspace{-0.3em}
\subsubsection{Synthetic Data Generation} 
FMs can also facilitate data generation, which enhances ST data by increasing its diversity, improving model robustness, and compensating for missing or sparse information \cite{long2024llms}. This is crucial for ST tasks like mobility analytics, where collecting real-world data is often costly or raises privacy concerns. For instance, Trajectory-LLM \cite{yang2025trajllm} generates vehicle trajectories from brief textual descriptions of vehicle interactions, whereas Traj-LLM~\cite{ju2025trajllm} generates human trajectories by leveraging personas, memory modules, and routine profiles. LLMob \cite{wang2024large} advances mobility data generation, offering flexibility in modeling diverse urban activities and personal mobility patterns, thus improving transportation system modeling and analysis. LLMs have also been employed to construct synthetic environments that replicate real-world conditions across diverse domains, including intelligent transportation \cite{adekanye2024llm} and disaster management \cite{goecks2023disasterresponsegpt}.

\subsection{Spatio-Temporal Data Management}

Upon the acquisition of ST data, the challenge of effective management emerges, particularly in addressing data quality issues (e.g., missing values/views) and facilitating data retrieval and integration. Within this context, FMs can be harnessed in the following ways.

\subsubsection{Data Cleaning} Data cleaning is the process of improving data quality by addressing issues such as missing values, low sampling rates, and noise. For example, ST data often exhibit missing values due to various factors like sensor malfunctions and transmission disruptions \cite{zheng2014urban}. Filling in these missing values\cite{qin2023multiple} is crucial for ensuring the integrity of predictive models, optimizing strategies, and facilitating informed decision-making. Recent literature reveals that LLMs can serve as powerful zero-shot \cite{zhang2024semantic} or few-shot \cite{cheng2024nuwats,zhang2024comprehensive} learners to data imputation by leveraging their ability to identify and learn complex ST patterns. PLMTrajRec \cite{wei2024ptr}, utilizing a pretrained language model to recover sparse trajectory data by unifying intervals and inferring road conditions, showing effective generalization across varied sampling intervals in tests. Moreover, scholars have investigated the potential of leveraging LLMs to augment missing views or information, such as urban region profiling \cite{yan2024urbanclip,yuan2024chatearthnet,hao2024urbanvlp} and traffic video captioning \cite{dinh2024trafficvlm}.

\vspace{-0.3em}
\subsubsection{Query \& Retrieval} Meanwhile, LLM can be applied to querying and retrieval to enhance information retrieval accuracy under the ST context. By leveraging their advanced natural language understanding capabilities, LLMs can process user queries in a more contextual and semantically rich manner, enabling precise retrieval of relevant information from structured and unstructured data sources. For instance, UrbanLLM \cite{jiang2024urbanllm} finetunes LLMs for urban activity planning and management, which serves as a problem solver that decodes urban-related queries into several sub-tasks, with each one solved by suitable spatio-temporal AI models. \citet{alamsyah2024automated} propose an automated smart city planning system that utilizes a personalized LLM with Retrieval Augmented Generation (RAG) \cite{gao2023retrieval} to generate tailored urban planning recommendations while ensuring data privacy, where RAG is used to retrieve relevant urban planning documents for context-aware responses. Another line of work \cite{li2024streetviewllm,zhong2024urbancross,zhangurbanmllm,licityanchor} utilizes Multimodal LLM for cross-modal information retrieval to enhance urban computing tasks.


\vspace{-0.3em}

\vspace{-0.3em}
\subsubsection{Data Integration} Data integration seeks to combine information from disparate sources, often necessitating the understanding and mapping of relationships between entities in heterogeneous datasets. LLMs are increasingly being employed in this domain, particularly for knowledge graph construction~\cite{ding2024zrllm}, where they automate and enhance the extraction, integration, and reasoning of related data. In the context of ST data, LLMs facilitate data integration by leveraging heterogeneous urban data sources, performing relational triplet extraction, and completing knowledge graphs through geospatial reasoning~\cite{ning2024uukg,liu2023urbankg}. A pioneering study UrbanKGent \cite{ning2024urbankgent} proposes an LLM-based Agent framework to automate the process of urban knowledge graph construction.  

\begin{table*}[t]
\caption{Summary of representative FMs tailored for ST data science.}\vspace{-1em}
\centering
\resizebox{\textwidth}{!}{
\begin{tabular}{>{\centering\arraybackslash}p{0.035\textwidth}  |
                >{\arraybackslash}p{0.25\textwidth} 
                >{\arraybackslash}p{0.45\textwidth}
                >{\centering\arraybackslash}p{0.2\textwidth}
                >{\centering\arraybackslash}p{0.07\textwidth}
                >{\centering\arraybackslash}p{0.15\textwidth}
                >{\centering\arraybackslash}p{0.07\textwidth}
                }
\shline
\textbf{Stage} &
  \textbf{Task \& Capability} &
  \textbf{Example} &
  \textbf{Method} &
  \textbf{Category} &
  \textbf{Venue} &
  \textbf{Year} \\\hline
\multirow{4}{*}{\textit{\rotatebox{90}{Sensing}}} &
  Real-World Data Sensing &
  Identifying Citizen-Related Issues from Social Media &
  dos Santos et al.~\cite{dos2024identifying} &
  LLM &
  CAiSE &
  2024 \\
 &
  Real-World Data Sensing &
  Intelligent Crowdsensing Coordination &
  AutoWebCrowds~\cite{zhu2024prototype} &
  LLM &
  ICWE &
  2024 \\
 &
  Synthetic Data Generation &
  Trajectories Generation &
  Trajectory-LLM~\cite{yang2025trajllm} &
  LLM &
  ICLR &
  2025 \\
 &
  Synthetic Data Generation &
  Human Activity Data Generation &
  LLMob~\cite{wang2024large} &
  LLM &
  NeurIPS &
  2024 \\\hline
\multirow{6}{*}{\textit{\rotatebox{90}{Management}}} &
  Data Cleaning &
  Few-Shot Learner for Filling Missing Values &
  NuwaTS~\cite{cheng2024nuwats} &
  PFM &
  Preprint &
  2024 \\
 
 &
  Data Cleaning &
  Trajectory Recovery & 
  PLMTrajRec~\cite{wei2024ptr} &
  LLM &
  Preprint &
  2024 \\
 &
  Data Cleaning &
  Augment Additional Views of Data &
  UrbanCLIP~\cite{yan2024urbanclip} &
  LLM &
  WWW &
  2024 \\
 &
  Query \& Retrieval &
  Autonomous Query Processor for Urban Management &
  UrbanLLM~\cite{jiang2024urbanllm} &
  LLM &
  EMNLP &
  2024 \\
 &
  Data Integration &
  Urban Knowledge Graph Construction &
  UrbanKGent~\cite{ning2024urbankgent} &
  LLM &
  NeurIPS &
  2024 \\\hline
\multirow{6}{*}{\textit{\rotatebox{90}{Mining}}} &
  Perception &
  Understand the Environment &
  Magma~\cite{yang2025magma} &
  PFM &
  CVPR &
  2025 \\
 &
  Perception &
  Interpret and Extract ST Patterns &
  STEP~\cite{shao2022pre} &
  PFM &
  KDD &
  2022 \\
 &
  Optimization &
  Drive Actionable Decision-Making in Dynamic Scenarios &
  AgentMove~\cite{feng2024agentmove} &
  LLM &
  Preprint &
  2024 \\
 &
  Optimization &
  Optimize Land-Use Plans by LLM Agents &
  Zhou et al.~\cite{zhou2024large} &
  LLM &
  Preprint &
  2024\\
 &
  Reasoning &
  Common-sense Reasoning &
  Causal-VidQA~\cite{li2022representation} &
  PFM &
  CVPR &
  2022 \\
 &
  Reasoning &
  Numerical Reasoning &
  UrbanGPT~\cite{li2024urbangpt} &
  LLM &
  KDD &
  2024 \\
 &
  Reasoning &
  Causal Reasoning &
  NuwaDynamics~\cite{wang2024nuwadynamics} &
  PFM &
  ICLR &
  2024 \\\hline
\multirow{7}{*}{\textit{\rotatebox{90}{Application}}} &
  Forecasting &
  Global Weather Forecasting &
  Pangu~\cite{bi2023accurate} &
  PFM &
  Nature &
  2023 \\
 &
  Imputation &
  Generative Adversarial Network for Traffic Data Imputation &
  STGAN~\cite{yuan2022stgan} &
  PFM &
  IEEE TBD &
  2022 \\
 &
  Anomaly Detection &
  Transformer-based Anomaly Detector &
  Xu et al.~\cite{xuanomaly} &
  PFM &
  ICLR &
  2022 \\
 &
  Event Analysis &
  Detecting and Interpreting Events &
  LAMP~\cite{shi2023language} &
  LLM &
  NeurIPS &
  2023 \\
 &
  Physical Grounding &
  Geo-localization &
  GeoGPT~\cite{zhang2023geogpt} &
  LLM &
  JAG &
  2023 \\
 &
  Decision Making &
  Transportation Analytics and Control &
  TrafficGPT~\cite{zhang2024trafficgpt} &
  LLM &
  Transport Policy  &
  2024 \\
 &
  Scenario Simulation &
  Simulation of Human Behavior &
  Park et al.~\cite{park2023generative} &
  LLM &
  UIST &
  2023\\

  \shline
\end{tabular}
}
\vspace{-1em}
\end{table*}

\subsection{Spatio-Temporal Data Mining}
Unlike traditional data mining, which primarily focuses on structured datasets, ST data mining captures intricate spatial and temporal dependencies within ST data using machine learning or deep learning techniques~\cite{jin2023spatio,wang2020deep,zhang2024survey}. With the emergence of FMs and LLMs, Spatio-Temporal Foundation Models (STFMs) offer new possibilities by integrating \emph{perception}, \emph{optimization}, and \emph{reasoning} capabilities to enhance ST data mining. In this section, we explore these key capabilities, while their specific applications across different domains are detailed in Sec. \ref{sec:app}.

\subsubsection{Perception}
In STFMs, perception encompasses the ability to effectively model, interpret, and generalize complex spatial and temporal patterns, enabling a deeper understanding of dynamic environments. This capability can be categorized into two key perspectives. The first view pertains to an \emph{agent}’s ability to perceive and understand its surrounding environment, capturing visual or contextual interactions within real-world scenarios such as smart cities \cite{yan2024opencity}, indoor activities \cite{yang2024thinking,yang2025magma}, and mobile Apps \cite{wang2024mobile}. 

The second aspect involves interpreting and extracting ST patterns from sensor data, ensuring accurate predictions across diverse domains. Domain-agnostic approaches, such as STEP~\cite{shao2022pre} and GPT-ST~\cite{li2024gpt}, have employed pretraining strategies that leverage historical observations to enhance forecasting performance. In urban computing, models like TFM~\cite{wang2023building} and OpenCity~\cite{li2024opencity} utilize graph-based FMs to analyze behaviors and interactions within transportation systems, yielding promising results in traffic prediction. In climate science, Pangu~\cite{bi2023accurate}, trained on 39 years of global climate data, delivers superior deterministic forecasting outcomes across all evaluated variables when compared to leading numerical weather prediction systems. Additional notable examples in this area include the works~\cite{li2024ocean,pathak2022fourcastnet, nguyen2023climax, lacoste2024geo}. Despite these advances, achieving robust generalization remains a critical challenge, as most existing research has been confined to in-domain applications. While models like UniST~\cite{yuan2024unist} are designed as one-for-all solutions for diverse ST scenarios, their training datasets and evaluation testbeds are predominantly limited to transportation. Nevertheless, their underlying technique stacks show promise for broader cross-domain and cross-modality generalization. Other significant contributions in this realm include UniFlow~\cite{yuan2024foundation} and UrbanDiT~\cite{yuan2024urbandit}.

\vspace{-0.65em}
\subsubsection{Optimization}
Building upon the perceptual foundations, the optimization ability focuses on refining and adapting models to achieve specific, task-oriented objectives. In other words, models are not only expected to capture rich ST patterns but also to drive actionable decision-making in dynamic, real-world scenarios. This involves integrating advanced optimization strategies that tailor model behavior to the unique demands of applications. 

A prominent approach involves \emph{agent-based frameworks}. For example, in traffic signal control, traditional methods (e.g., RL) are now augmented by frameworks that use LLMs as decision-making agents~\cite{lai2023large}. These systems leverage real-time traffic data and expert prompts to enable human-like planning, resulting in more adaptive and interpretable control strategies. Similarly, CityGPT~\cite{guan2024citygpt} decomposes ST analysis into specialized sub-tasks, handled by temporal, spatial, and fusion agents, to efficiently process IoT data and generate insightful visualizations. AgentMove~\cite{feng2024agentmove} addresses human mobility prediction by breaking down the task into modules for individual pattern mining, urban structure analysis, and collective behavior extraction. In geo-science, systems like Geode~\cite{gupta2024geode} integrate explicit optimization modules with ST data retrieval and machine learning inference to tackle zero-shot geospatial QA with enhanced precision. In urban planning, an innovative work \cite{zhou2024large} simulates planners and residents by LLM agents and enables their interactions to optimize inclusive land-use plans efficiently. Despite these promising developments, significant challenges remain. Seamlessly integrating perceptual capabilities with targeted optimization strategies is crucial for next-generation ST models that are both versatile and effective across diverse operational contexts.

\subsubsection{Reasoning}
While current ST models have demonstrated notable success in recognition and agent-based tasks, their reasoning and cognitive capabilities remain underdeveloped compared to advanced systems like DeepSeek-R1~\cite{guo2025deepseek}. To progress toward ST general intelligence, we identify three key aspects of reasoning: 
\vspace{-0.2em}
\begin{itemize}[leftmargin=*]
    \item \textbf{Common-sense Reasoning} harnesses everyday knowledge and contextual cues to draw implicit inferences from complex data. For instance, Causal-VidQA~\cite{li2022representation}enables models to infer explanations, predict future states, and generate counterfactual scenarios in video question-answering, while SituatedGen~\cite{zhang2024situatedgen} integrates geographical and temporal contexts to generate coherent and contextually plausible statements.

    \item \textbf{Numerical Reasoning} involves interpreting and manipulating quantitative information to perform arithmetic operations, assess uncertainties, and discern relationships within ST data; for instance, STBench~\cite{li2024stbench} evaluates these abilities in LLMs, while UrbanGPT~\cite{li2024urbangpt} enhances ST forecasting with instruction tuning. 

    \item \textbf{Causal Reasoning} seeks to uncover cause-effect relations within ST data, crucial for robust and interpretable predictions. For example, NuwaDynamics~\cite{wang2024nuwadynamics} identifies causal regions and applies interventions to improve generalization, and GCIM~\cite{zhao2023generative} learns latent causal structures to disentangle spurious correlations. 
\end{itemize}
\vspace{-0.2em}
Collectively, these dimensions offer a promising yet underexplored pathway toward achieving ST general intelligence, bridging the gap between pattern recognition and true cognitive understanding.

\vspace{-0.5em}
\subsection{Downstream Applications} \label{sec:app}
\subsubsection{STFMs for Numerical Problems}
ST data is predominately numeric in many real-world scenarios. Addressing these numeric challenges is critical for tasks like forecasting, imputation, and anomaly detection~\cite{jin2024survey}, which demand an accurate understanding of the physical world. STFMs excel in these areas by uncovering intricate patterns and dependencies, ultimately enabling more reliable data-driven decision-making.

\textbf{\textit{- Forecasting.}} Early forecasting approaches often relied on task-specific neural networks like STGNNs~\cite{jin2023spatio,jin2024survey,prabowo2023because,shao2022long}, whereas recent developments have shifted toward universal forecasting~\cite{liu2024unitime,woo2024unified,zhang2024survey}. For instance, GPT-ST~\cite{li2024gpt} leverages pretraining on historical observations to boost predictive performance, while UniST~\cite{  yuan2024unist} unifies multiple traffic prediction tasks within a single model by coupling sequence modeling with attention-based mechanisms. Building on this progress, ST-LLM~\cite{liu2024spatial} and STG-LLM~\cite{liu2024can} enhance traffic predictions by combining ST inputs with partially frozen large language models, and UrbanGPT~\cite{li2024urbangpt} extends this paradigm further by employing ST instruction tuning to better align textual and ST data. Similar approaches have also been widely used in other domains, such as ClimaX~\cite{nguyen2023climax}, Geo-Bench~\cite{lacoste2024geo}, and Orca~\cite{li2024ocean}.

\textbf{\textit{- Imputation.}} This has likewise benefited from techniques that capture ST dependencies to accurately restore missing or corrupted data. For instance, NuwaTS~\cite{cheng2024nuwats} repurposes pretrained language models with contrastive learning and specialized patch embeddings (capturing missing patterns/statistics) to enable cross-domain time series imputation through a unified framework. STD-LLM~\cite{huang2024std} employs LLMs with spatial-temporal tokenizers and hypergraph learning modules to handle missing values in spatio-temporal data while capturing non-pairwise correlations through topology-aware node embeddings. DrIM~\cite{limcontext} combines LLM-derived text representations (from masked tabular data conversions) with contrastive learning to measure similarities for nearest-neighbor imputation in heterogeneous datasets.

\textbf{\textit{- Anomaly Detection.}} Anomaly detection in ST data has advanced by leveraging models that learn the normal dynamics of ST systems to identify deviations indicative of abnormal events. Whereas prior methods relied on statistical thresholding and clustering to flag outliers, recent FMs learn robust ST representations to detect even subtle anomalies. For example, early attempts \cite{dong2024can,liu2024large,zhou2024can} investigate the feasibility of using LLMs for anomaly detection in time series data. SigLLM~\cite{alnegheimish2024large} employs GPT-series with signal-to-text conversion techniques, offering dual pipelines (anomaly prompting and deviation detection) for time series analysis through textual or visual representations of numerical data.  AD-LLM \cite{yang2024ad} introduces a benchmark framework combining GPT-4's zero-shot reasoning with contrastive learning for anomaly context enrichment and automated model selection through chain-of-thought prompting.

\textbf{\textit{- Others.}} Furthermore, FMs have demonstrated great potential in other numerical problems such as time series classification \cite{cheng2024advancing}, geospatial prediction \cite{manvi2024geollm,gurnee2023language}, traffic speed inference \cite{balsebre2024city}, and socio-economic indicator prediction \cite{hao2024urbanvlp,yan2024urbanclip,xiao2024refound}.


\subsubsection{STFMs for Inferential Problems}
Inferential problems in ST data require the integration of both reasoning and understanding of environments. These problems involve high-level cognitive tasks where accurate representation of locations, movements, and environmental context is essential. Addressing such problems goes beyond numerical predictions — it necessitates answering critical inferential questions: \textit{What happened? Where is it? What to do? What if?}
FMs have shown their potential to enhance solutions for these challenges by leveraging their capacity to handle ST knowledge and interpret complex, unstructured data.

\vspace{-0.5em}
\paragraph{\textbf{\textit{``What happened?'' - Event Analysis}}} Detecting events aims to recognize and explain significant events in time and space. Traditional models struggle with scalability, interpretability, and incorporating external knowledge. To this end, LAMP \cite{shi2023language} integrates LLMs with event models, using abductive reasoning to suggest plausible causes for predicted events, retrieve supporting evidence, and rank predictions for improved accuracy. Meanwhile, LEAP \cite{zhang2024large} replaces GNNs and RNNs with LLMs by framing event detection as a question-answering task, predicting missing event components and forecasting future relations through self-attention mechanisms.

\vspace{-0.5em}
\paragraph{\textbf{\textit{``Where is it?''- Physical Grounding.}} }
Grounding ST models in real-world geographical contexts is essential for various applications such as geo-localization, map reconstruction, intelligent routing and navigation.  
Geo-localization aims to determine an object's location based on multimodal inputs, including images, text, and sensor data. By processing these cues in conjunction with map data, LLMs such as GPT-4o, DeepSeek~\cite{guo2025deepseek}, and GeoGPT~\cite{zhang2023geogpt} can infer geographic coordinates or identify specific locations described in natural language. 
Map reconstruction, on the other hand, involves creating or updating digital maps by synthesizing information from satellite imagery, sensor readings, and textual reports. LLMs contribute by interpreting and generating map content, correcting inaccuracies, and filling in missing details. For instance, MapGPT~\cite{chen2024mapgpt} employs language-guided updates, incorporating textual descriptions of environmental changes into existing map structures. 
In personalized routing, ItiNera~\cite{tang2024itinera} combines LLMs with spatial optimization to generate personalized ``Citywalk'' itineraries, providing user-specific and spatially coherent urban exploration; ChinaTravel~\cite{shao2024chinatravel} provides a benchmark for real-world Chinese travel planning, enabling scalable evaluation of constraint satisfaction and preference optimization while highlighting the strengths of neuro-symbolic agents.
Navigation systems further benefit from LLMs' ability to understand contextual instructions, interpret user queries, and reason about dynamic environments. For example, NavGPT~\cite{zhou2024navgpt} and NavGPT-v2~\cite{zhou2024navgpt2} integrate natural language with real-time traffic and indoor video data to generate personalized and optimized routing solutions.
By incorporating STFMs across these domains, physical grounding models facilitate more precise localization, efficient navigation, and adaptive urban mobility solutions, ultimately bridging the gap between digital intelligence and real-world spatial reasoning.  

\vspace{-0.5em}
\paragraph{\textbf{\textit{``What to do?'' - Decision Making.}} }
Optimizing policies and real-time decision-making in dynamic environments based on inferential insights plays a crucial role in a wide range of applications, including traffic control, autonomous vehicles, and disaster response.
In traffic control and management, LLMs improve adaptability and interpretability compared to traditional reinforcement learning approaches~\cite{lai2023large}. Additionally, they facilitate sim-to-real transfer by modeling real-world traffic dynamics, improving the reliability of traffic signal optimization~\cite{da2023llm}. Beyond signal control, models like TrafficGPT~\cite{zhang2024trafficgpt} integrate multimodal traffic data with structured reasoning to analyze, predict, and optimize traffic efficiency and safety in real time.  
In autonomous vehicles, STFMs contribute to decision-making through both direct and indirect mechanisms. Directly, models such as DDM-Lag~\cite{liu2024ddm} employ diffusion-based frameworks with Lagrangian safety enhancements and hybrid policy updates to refine policy articulation and ensure safety. Indirectly, STFMs enhance autonomous driving by predicting realistic driving behaviors~\cite{jin2023surrealdriver,sha2023languagempc} and leveraging multi-modal perception to integrate sensor data, bird’s eye view maps, and traffic contexts~\cite{choudhary2024talk2bev,zhou2023vision}, improving situational awareness and vehicle control.
Beyond transportation, STFMs play a critical role in disaster management and emergency response by integrating diverse spatio-temporal data sources, such as weather forecasts, remote sensing, and social media signals, to predict disaster impacts and optimize evacuation strategies~\cite{goecks2023disasterresponsegpt,chen2024enhancing,lei2025harnessing}.  

\begin{figure}[!t]
    \centering
    \includegraphics[width=0.99 \linewidth]{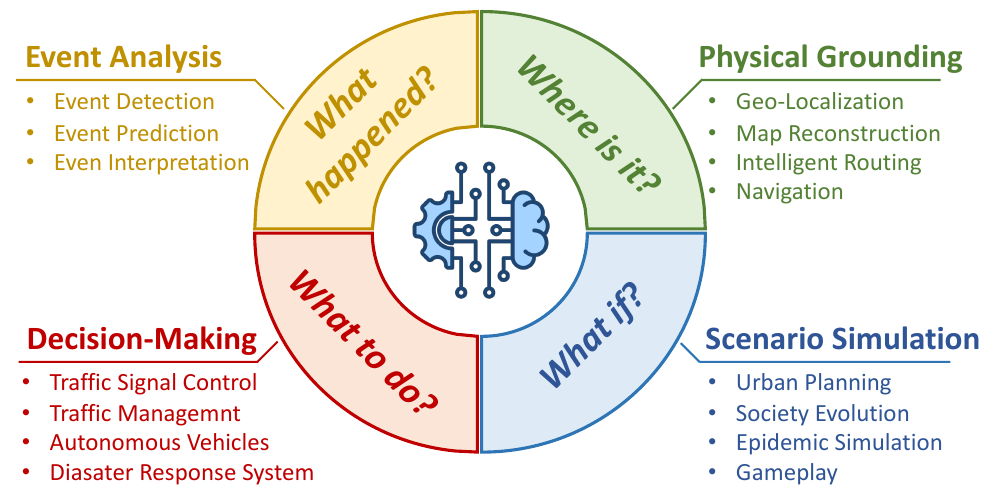}
    \vspace{-0.5em}
    \caption{STFMs for addressing inferential problems.}
    \vspace{-1.5em}
    \label{fig:inferential_task}
\end{figure}

\vspace{-0.5em}
\paragraph{\textbf{\textit{``What if?'' - Scenario Simulation.}}}  
STFMs, with their advanced perception and reasoning capabilities, enable the development of STFM-based agents that integrate into Multi-Agent Systems (MAS) to model complex interactions across diverse domains~\cite{gao2024large}. 
In urban planning and social simulation, MAS facilitates participatory urban design by simulating interactions between planners and residents. For example, LLM-driven MAS has been used to collaboratively refine land-use plans, leading to improved accessibility and ecological outcomes that surpass human expert solutions~\cite{zhou2024large}. Beyond urban planning, MAS contributes to social science research by modeling human-like behaviors in AI-driven networks. Studies such as~\cite{park2023generative,dai2024artificial,piao2025emergence} demonstrate that LLM-based agents can naturally develop social structures, providing valuable insights into emergent social dynamics.  
Beyond urban applications, MAS significantly advances game AI and strategic decision-making. Recent studies~\cite{wang2023describe,zhu2023ghost,qicivrealm} highlight how MAS-powered reinforcement learning enables strategic gameplay, real-time opponent modeling, and interactive storytelling, fostering the development of more adaptive, intelligent, and realistic virtual agents.

\section{The Methodology Perspective} 
As shown in Figure~\ref{fig:taxonomy}, we delve into STFMs from a methodology perspective, focusing on $i$) LLM-based models, which are widely applied \textit{across the entire workflow of ST data science} by zero-shot utilization or fine-tuning and $ii$) PFM-based models, i.e., pretraining FMs from scratch, which is \textit{mainly utilized for ST data mining.} The comparison between them can be found in Appendix \ref{appendix:comparison}.
\input{img/taxonomy}

\subsection{Large Language Models (LLM)} \label{sec:LLM}
\subsubsection{Zero-shot Learner} LLMs exhibit strong reasoning and contextual understanding capabilities, making them highly effective across various ST tasks, including data sensing, management, and mining. As shown in Appendix~\ref{appendix:llm-usage}, they can function as augmenters, predictors, or agents. To ease the presentation, we adopt a broad definition of LLMs, encompassing standard LLMs, Vision-Language Models (VLM), and Multimodal LLMs (MLLM). The zero-shot utilization of LLMs can be categorized into two primary classes.

\vspace{0.2em}
\textbf{- \textit{Prompt Engineering}}. When taking LLMs as zero-shot predictors \cite{jin2023timellm,gruver2023large,vaghefi2023chatclimate} or data augmenters \cite{yan2024urbanclip} for various tasks, prompt engineering plays an essential role in shaping model outputs. Below, we summarize key aspects for prompt engineering in current research: 
a) \textbf{{Prompt Construction:}} A well-designed prompt typically contains key elements like \emph{Task Instruction}, \emph{Tokenization}, and \emph{Few-shot Examples}. Task instruction~\cite{jin2023timellm,xue2022leveraging,xue2023promptcast} aims to explicitly guide LLMs to execute specific operations, incorporating domain knowledge~\cite{yu2023temporal} if applicable. Tokenization \cite{jin2023timellm,gruver2023large} is crucial to aligning ST data formats with LLM input structures. Additionally, presenting a small number of annotated examples~\cite{zhang2024llm4dyg} facilitates in-context learning, enabling LLMs to better generalize to complex tasks while ensuring output consistency and adherence to the expected format.
b) \textbf{{Prompt Learning:}} \cite{li2024urbangpt,xue2024prompt} Also known as instruction-tuning, this method learns prompts dynamically rather than relying on manually crafted ones. By optimizing prompt structures during training, it provides a flexible and efficient way to adapt LLMs to new tasks without altering their underlying model weights.
c) \textbf{{Chain-of-Thought (CoT) Prompting:}} CoT~\cite{liu2024lstprompt,zhang2024llm4dyg} enhances LLMs' reasoning capabilities by guiding them through step-by-step logical progression. This method improves their ability to tackle complex spatio-temporal tasks, ensuring more interpretable, structured, and accurate outputs in decision-making processes.




\vspace{0.2em}
\par \textbf{- \textit{Agentic Engineering}}. The emergence of LLM-based agents~\cite{zhang2024trafficgpt,xu2023urban,zhang2023geogpt,zhou2024large,jiang2025explainable} with reasoning, memory and tool-calling capabilities is transforming ST data science, enabling more adaptive and autonomous decision-making. When designing agent-based solutions, existing works primarily consider the following key aspects: 
a) \textbf{{Role Assignment.}}~\cite{xu2023urban,zhang2023geogpt,jiawei2024large} clearly specify the responsibilities and functional boundaries of each agent within the system. 
b) \textbf{{Memorization}}~\cite{zhang2023geogpt,lei2025stma} refers to the agent’s capability to store, recall, and leverage past information and context during task execution. 
A basic approach involves embedding past interactions into prompts, while more advanced techniques like Retrieval-Augmented Generation (RAG) \cite{yang2024timerag,xiao2025retrieval} dynamically retrieve relevant information from external knowledge bases, incorporating only the most pertinent content into the prompt.
c) \textbf{{Tool Definition}}~\cite{zhang2024trafficgpt,zhang2023geogpt}, which identify and integrate various tools and functionalities that an agent can call upon to solve complex tasks. In ST data science, various expert models like STGNNs \cite{jin2023spatio} can be wrapped as a tool and added into the agent in a plug-and-play manner.
d) \textbf{{Multi-Agent System.}} Deploying multiple specialized agents to work collaboratively (each with distinct roles) enhances the efficiency and robustness of solutions for intricate ST challenges \cite{zhou2024large,jiang2025explainable,lee2025timecap}.

\subsubsection{Supervised Fine-Tuning for LLMs}
Fine-tuning adapts LLMs to ST tasks by adjusting their parameters based on domain-specific datasets, sometimes incorporating additional modalities such as texts~\cite{liang2023exploring,yan2024urbanclip} and vision~\cite{zhong2025time}. We categorize fine-tuning methods into three approaches based on \emph{the extent of parameter updates}:
\begin{itemize}[leftmargin=*]
    \item \textbf{Full Parameter Fine-Tuning}~\cite{man2023w,nguyen2023climax,pathak2022fourcastnet,manvi2024geollm,li2024large} updates all model parameters based on downstream ST datasets, achieving maximal adaptation to specific tasks. However, it requires substantial labeled data and high computational resources, making it impractical for many real-world applications.

    \item \textbf{Partial Parameter Fine-tuning.} To reduce computational overhead, this method~\cite{chang2023llm4ts,zhou2023one} freezes most parameters, such as attention weights, while fine-tuning only a small subset (e.g., position encodings and layer normalization). However, modifying a subset of parameters can disrupt the LLM’s learned representations, leading to catastrophic forgetting of general knowledge.

    \item \textbf{Add-on Parameter Fine-Tuning.} To mitigate forgetting while maintaining efficiency, this technique~\cite{lai2023large} introduces trainable low-rank matrices (e.g., LoRA~\cite{hu2021lora}), while keeping the original LLM weights frozen. This strategy preserves pretrained knowledge while enabling efficient adaptation to ST tasks. Besides fine-tuning LLMs' weights, another way is training additional layers for input tokenization or task adaption. For instance, Time-LLM~\cite{jin2023timellm} trains a self-attention layer that aligns patched time series representations with pretrained text prototype embeddings. Similarly, Time-VLM~\cite{zhong2025time} trains a memory-enhanced attention to capture both short- and long-term dependencies. For task adaption, existing methods typically train an additional prediction head (e.g., linear layers) to project the LLM's output embeddings into a domain-specific space \cite{jin2023timellm,zhong2025time}.
\end{itemize}


\subsection{Pretrained Foundation Models (PFM)} \label{sec:PFM}

Unlike LLMs, which build STFMs by directly using or fine-tuning LLMs, PFMs are developed from scratch, independent of existing LLMs. This approach enables domain-specific optimization, allowing models to better capture ST dependencies from cross-domain ST data without constraints imposed by linguistic priors. Following this, we examine PFMs through three key dimensions:
\vspace{-0.3em}
\subsubsection{Neural Architecture} The architecture of PFMs is a fundamental design choice that directly influences their capabilities, efficiency, and adaptability in ST tasks, which can be categorized into:
\begin{itemize}[leftmargin=*]
    \item \textbf{Transformer-based PFMs}. Transformers have been the predominant architecture choice for building PFMs thanks to its powerful sequential modeling ability introduced by the self-attention mechanism \cite{lin2024trajfm,li2023geolm,balsebre2024city,li2024opencity,yuan2024unist}.  

    \item \textbf{Diffusion-based PFMs}. Diffusion-based models have recently emerged as a powerful approach for ST representation learning \cite{cao2024timedit,yuan2024urbandit,zhu2024controltraj,yuan2024spatio,chu2023trajgdm,wen2023diffstg}, particularly in generative and predictive modeling. These models iteratively learn to reverse a stochastic noise process, enabling them to generate high-fidelity spatio-temporal sequences with strong generalization properties.

    \item \textbf{Graph-based PFMs}. Unlike sequential models, GNNs excel at representing spatially structured data such as road networks. \cite{wang2023building,lam2023learning} build FMs based on graph neural networks to learn the complex correlation between different entities in ST applications.

    \item \textbf{Others}. Another emerging class of PFMs is State Space Model (SSM)-based models \cite{ma2024mamba,bhethanabhotla2024mamba4cast,hu2024time}, which construct PFMs using structured state-space representations. Meanwhile, several studies utilize CNNs~\cite{dada} as backbones for developing PFMs.
\end{itemize}

\vspace{-0.3em}
\subsubsection{Pretraining Scheme} To enhance generalization ability, PFMs are usually pretrained based on cross-domain datasets \cite{liu2024unitime,woo2024unified,yuan2024unist}, enabling them to learn diverse ST patterns across multiple domains. Existing pretraining schemes of PFMs can be classified into three types based on the training objectives:
a) \textbf{Generative Pretraining} \cite{lin2024trajfm,wang2023building,man2023w,wu2024g2ptl,zhu2024unitraj} focuses on reconstructing input data by learning its underlying distribution, enabling the model to generate realistic time series or ST data, while b) \textbf{Contrastive Pretraining} \cite{balsebre2024city,lin2024ptrajm,zhang2024urban} emphasize distinguishing between similar and dissimilar data pairs to learn robust representations by maximizing agreement between augmented views of the same sample. It is particularly effective in multimodal ST learning, aligning heterogeneous data sources such as satellite imagery and its text description. 
c) \textbf{Hybrid Pretraining} \cite{li2023geolm} integrates both generative and contrastive objectives, leveraging their complementary strengths.

\vspace{-0.3em}
\subsubsection{Data Modality} 
ST data manifests in various modalities, each characterized by unique properties (see Section~\ref{sec:STD_data}), necessitating the development of modality-specific STFMs:

\vspace{-0.1em}
\begin{itemize}[leftmargin=*]
    \item \textbf{Location.} PFMs for location data~\cite{balsebre2024city,tempelmeier2021geovectors,li2022spabert,wu2024g2ptl,zhang2024urban,yan2024urbanclip,hao2024urbanvlp} aim to learn general embedding for geographical entities. For instance, GeoVectors \cite{tempelmeier2021geovectors} and SpaBERT~\cite{li2022spabert} learn location embeddings based on open-source data such OpenStreetMap, while G2PTL~\cite{wu2024g2ptl} learns from massive logistics delivery data. Notably, there is a noticeable trend that leverages multi-modalities (such as satellite image and text) for comprehensive location embeddings. For example, both UrbanCLIP~\cite{yan2024urbanclip}, UrbanVLP~\cite{hao2024urbanvlp}, and ReFound~\cite{xiao2024refound} utilize satellite images for urban region profiling.

    \item \textbf{Trajectory \& Event.} PFMs for trajectory/event data  \cite{lin2024trajfm,najjar2023towards,lin2024ptrajm,zhu2024unitraj,chu2023trajgdm,steinberg2023motor} are designed to learn general sequential patterns from inputs. A pioneering effort in this direction is TrajFM~\cite{lin2024trajfm}, which introduces a trajectory FM capable of supporting both regional and task transferability. Pretrained on vehicle trajectories from multiple cities, TrajFM employs a trajectory-masking and autoregressive recovery mechanism to enhance its learning capabilities. To tackle the limited resources of cross-domain trajectories, UniTraj~\cite{zhu2024unitraj} curates a billion-scale mobility dataset spanning diverse geographic regions to facilitate the advancement of trajectory-based FMs. For event data, MOTOR~\cite{steinberg2023motor} proposes a time-to-event FM for structured medical records.

     \item \textbf{ST Raster.} PFMs for ST raster data \cite{man2023w,yuan2024urbandit,nguyen2023climax,bi2022pangu,pathak2022fourcastnet,shao2022pre,chen2023fengwu} organize spatial information in a grid-like format, with a typical applied domain being weather/climate forecasting. For instance,  W-MAE~\cite{man2023w} trains a mask autoencoder for ST grid forecasting. ClimaX~\cite{nguyen2023climax} develops a general-purpose climate foundation model, pretrained on diverse datasets spanning various variables, ST scales, and physical contexts. Pangu~\cite{bi2022pangu} is trained on 39 years of global climate data, which achieves superior forecasting performance compared to leading numerical weather prediction systems. UniST~\cite{yuan2024unist} first pretrains the model in various ST raster data via masked pretraining, and then proposes a learnable ST prompt to enhance the model's generalization ability.

    \item  \textbf{ST Graph.} PFMs for ST graph data~\cite{li2024opencity,shao2022pre,lam2023learning,prabowo2024traffic,wang2022event,liu2024moirai} learn the ST dependencies from ST graphs that generalize effectively in unseen spatial and temporal contexts. Unlike ST Raster PFMs, there are limited works in this area, which is more challenging due to the complex graph correlation. One representative is OpenCity~\cite{li2024opencity} for ST graph forecasting, which integrates Transformer and GNN to model the ST dependencies in traffic data.

\end{itemize}
\begin{figure}[!t]
    \centering
    \includegraphics[width=0.99 \linewidth]{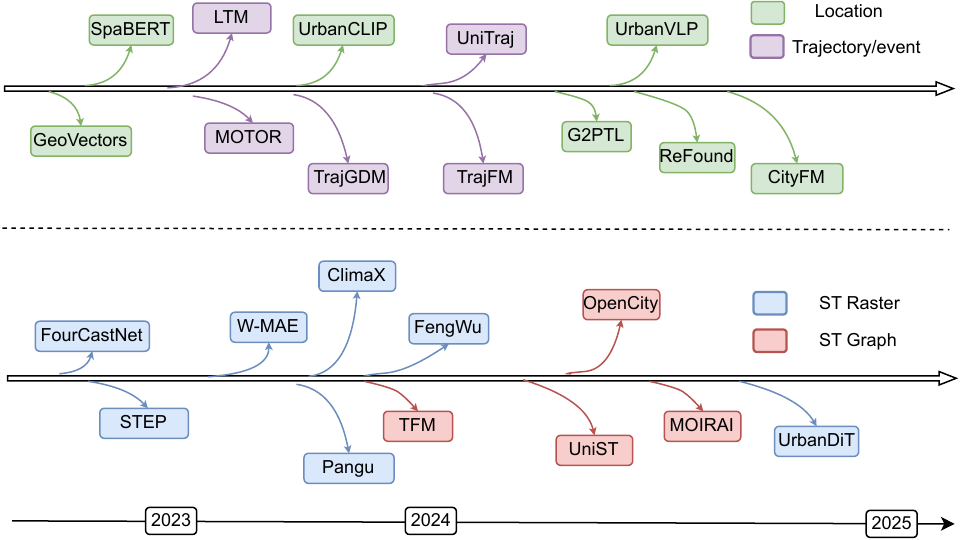}
    \vspace{-1em}
    \caption{Representative PFMs for different types of ST data.}
    \vspace{-1.5em}
    \label{fig:data_specific_models}
\end{figure}

\vspace{-1em}
\section{Conclusion and Future Directions} 
\vspace{-0.2em}
The rapid advancement of FMs has transformed ST data science, impacting sensing, management, and mining. This survey provides a comprehensive review of FMs for ST data science, identifying key capabilities such as perception, reasoning, and optimization while exploring diverse downstream tasks and datasets. We also establish a systematic taxonomy of methodologies, enhancing understanding of how STFMs model ST data. Despite progress, challenges remain in generalization, interpretability, and efficiency. By consolidating recent advances and outlining future directions (see Appendix \ref{sec:future}), this survey aims to inspire further innovations, driving the development of scalable and adaptive STFMs for real practice.


\balance
\bibliographystyle{ACM-Reference-Format}
\bibliography{ref}

\appendix
\balance

\input{img/taxonomy_full}

\section{Limitations and Future Opportunities} \label{sec:future}
We further discuss the potential limitations of current research and identify several key future directions aimed at advancing the development of more powerful, transparent, and reliable STFMs:
\begin{itemize}[leftmargin=*]
    \item \textbf{The curse of accuracy against interpretability.} We have identified a significant challenge in developing FMs for addressing numerical problems in ST data science. Directly leveraging LLMs for numerical tasks such as forecasting proves to be non-trivial \cite{gruver2024large}. Meanwhile, fine-tuning LLMs or training STFMs from scratch using large-scale, cross-domain ST data often comes at the cost of interactive capabilities, thereby hindering interpretability in the prediction outcomes. These limitations motivate us to explore a novel paradigm that not only retains strong numerical reasoning abilities but also enhances interpretability, bridging the gap between predictive accuracy and explanatory insight.

    \item \textbf{Large foundation models are all we need?} While the extensive parameterization of FMs enables impressive generalization capabilities, particularly in zero/few-shot settings, their superiority over smaller expert models remains context-dependent. In ST domains such as time series analysis \cite{tan2024language} and urban planning \cite{kambhampati2024llms}, smaller expert models often outperform FMs when provided with sufficient domain-specific training data. This raises fundamental questions about the \emph{trade-offs} between model scalability, efficiency, and task-specific optimization. Future research should delve into hybrid approaches that combine the adaptability of large models with the precision of expert models.

    \item \textbf{One-fit-all FMs across the full workflow.} While current FMs are typically designed to support only specific stages of ST data science, we envision a more unified FM capable of seamlessly spanning the entire workflow, from initial data sensing and management to mining and supporting downstream applications. Achieving this goal will likely require the development of advanced LLM agents that can function as full-stack engineers (i.e., strongly benefiting all stages) for ST data science.
    
   \item \textbf{Integrating STFMs with multimodal understanding.} While current STFMs excel in processing structured ST data, their ability to integrate and reason over multimodal information, including text, images, video, and sensor data, remains underdeveloped. Many tasks require models to jointly interpret geospatial context, temporal dynamics, and text descriptions. Future research can focus on designing multimodal STFMs that effectively align, fuse, and reason over heterogeneous data sources, enabling more context-aware and human-interpretable decision-making.
    
\end{itemize}

\section{Zero-shot Utilization of LLMs} \label{appendix:llm-usage}

There are three ways of directly using LLMs for various ST tasks: 
\begin{itemize}[leftmargin=*]
    \item \textbf{LLM-as-Augmenter.} Pretrained LLMs can enhance both data understanding and model performance. On the one hand, it can serve as the input augmenter, which enhances data interoperability or provides external information \cite{hao2024urbanvlp,liang2023exploring} (e.g., textual or visual).  On the other hand, LLMs can serve as a parameter-frozen model component \cite{anonymous2024st, moon2023imu2clip, yan2024urbanclip}, thus augmenting domain models by injecting the pretrained external knowledge in LLMs. 

    \item \textbf{LLM-as-Predictor.}  LLMs can be directly employed as predictors~\cite{jin2023timellm,gruver2023large,li2024urbangpt,vaghefi2023chatclimate} for various tasks. Due to the modality gap between text and ST data, preprocessing is required to fit the input spaces of LLMs. Such step typically contains prompt engineering~\cite{vaghefi2023chatclimate,li2024urbangpt,xue2022leveraging,xue2023promptcast,xue2024prompt} or patch \& tokenization~\cite{jin2023timellm}.

    \item \textbf{LLM-as-Agent.} LLM-based agents are typically equipped with the ability to memorize and call various tools. When applied to ST data science, various domain-expert models can be wrapped as a tool and added into the agent in a plug-and-play manner \cite{zhang2024trafficgpt, xu2023urban,zhang2023geogpt}. As such, LLM serves as a router to access different models with both flexibility and performance guarantees. Furthermore, multi-agent systems \cite{zhou2024large} can be built to solve more complex tasks in the ST domain.
\end{itemize}

\section{Comparison between LLMs and PFMs} \label{appendix:comparison}
Table \ref{tab:llm_pfm} demonstrates the comparison between LLMs and PFMs on their capabilities, including perception, optimization, and reasoning. For example, PFMs excel in exceptional numerical reasoning abilities, yet they often struggle with common-sense understanding. There is still no free lunch, and the user can choose either LLMs or PFMs according to the downstream applications.

\begin{table}[!h]
\caption{A capability comparison between LLMs and PFMs for ST data science. }
\centering
\resizebox{0.47\textwidth}{!}{
\begin{tabular}{>{\centering\arraybackslash}p{0.12\textwidth}  |
                >{\arraybackslash}p{0.3\textwidth} |
                >{\arraybackslash}p{0.3\textwidth}
                }
                \shline
\textbf{Capabilities}  & \textbf{Large Language Models (LLMs)}                         & \textbf{Pretrained Foundation Models (PFMs)}                                \\\hline
Perception &
  \warn  Limited native ST perception; can be enhanced via fine-tuning &
  \cmark Strong ST perception, integrating sensor data and domain-specific learning \\\hline
Optimization &
  \cmark Agent-based reasoning for decision-making; relies on prompting and heuristics &
  \warn Limited; lacks decision-making ability for control and planning \\\hline
Common-sense Reasoning & \cmark Strong via pretraining on vast textual data; can be enhanced with fine-tuning & \warn Limited; relies on structured ST data rather than broad world knowledge \\\hline
Numerical Reasoning &
  \warn Handles arithmetic but struggles with structured ST computations &
  \cmark Designed for numerical problems, e.g., forecasting, anomaly detection \\\hline
Causal Reasoning &
  \warn Can infer causal relations from text but lacks structured ST modeling &
  \cmark Built-in graph-based and ST causal modeling \\
  \shline
\end{tabular}
}
\label{tab:llm_pfm}
\end{table}

\clearpage
\end{document}

%% file: img/taxonomy.tex
\tikzstyle{my-box}=[
    rectangle,
    draw=hidden-draw,
    rounded corners,
    text opacity=1,
    minimum height=1.5em,
    minimum width=5em,
    inner sep=2pt,
    align=center,
    fill opacity=.5,
    line width=0.8pt,
]
\tikzstyle{leaf}=[my-box, minimum height=1.5em,
    fill=hidden-pink!80, text=black, align=left, font=\normalsize,
    inner xsep=2pt,
    inner ysep=4pt,
    line width=0.8pt,
]
\begin{figure}[!h]
    \centering
    \vspace{-0.5em}
    \resizebox{0.47 \textwidth}{!}
    {
        \begin{forest}
            forked edges,
            for tree={
                fill=level0!80,
                grow=east,
                reversed=true,
                anchor=base west,
                parent anchor=east,
                child anchor=west,
                base=left,
                font=\large,
                rectangle,
                draw=hidden-draw,
                rounded corners,
                align=left,
                minimum width=4em,
                edge+={darkgray, line width=1pt},
                s sep=3pt,
                inner xsep=2pt,
                inner ysep=3pt,
                line width=0.8pt,
                ver/.style={rotate=90, child anchor=north, parent anchor=south, anchor=center},
            },
            where level=1{text width=2em,font=\normalsize,fill=level1!80,}{},
            where level=2{text width=9em,font=\normalsize,fill=level2!80,}{},
            where level=3{text width=9em,font=\normalsize,fill=level3!60,}{},
            where level=4{text width=10em,font=\normalsize,fill=level4!20,}{},
            where level=5{text width=5em,font=\normalsize,}{},
            [STFM  
                [LLM
                    [Zero-shot Learner
                        [Prompt Engineering
                            [\cite{jin2023timellm,xue2022leveraging,xue2023promptcast,gruver2023large,zhang2024llm4dyg,liu2024lstprompt}, leaf, text width=12em ]
                        ]
                        [Agentic Engineering
                            [\cite{zhang2023geogpt,jiawei2024large,lei2025stma,yang2024timerag,zhang2024trafficgpt,jiang2025explainable,lee2025timecap}, leaf, text width=12em]
                        ]
                    ]
                    [Fine-tuning
                        [All Parameters
                            [\cite{man2023w,nguyen2023climax,pathak2022fourcastnet,manvi2023geollm}, leaf, text width=12em ]
                        ]
                        [Partial Parameters
                            [\cite{chang2023llm4ts,zhou2023one,liu2024spatial}, leaf, text width=12em ]
                        ]
                        [Add-on Parameters
                            [\cite{jin2023timellm,lai2023large,zhong2025time} , leaf, text width=12em ]
                        ]
                    ]
                ]
                [PFM
                    [Neural Architecture
                        [Transformer-based
                            [\cite{lin2024trajfm,li2023geolm,balsebre2024city,li2024opencity,yuan2024unist,rose}, leaf, text width=12em ]
                        ] 
                        [Diffusion-based
                            [\cite{cao2024timedit,yuan2024urbandit,yuan2024spatio,chu2023trajgdm,long2024universal}, leaf, text width=12em ]
                        ]
                        [Graph-based
                            [\cite{wang2023building,lam2023learning}, leaf, text width=12em ]
                        ]
                        [Others 
                            [\cite{ma2024mamba,bhethanabhotla2024mamba4cast,hu2024time,kochkov2024neural,dada}, leaf, text width=12em ]
                        ]
                    ]
                    [Pre-training Scheme
                        [Generative
                            [\cite{lin2024trajfm,wang2023building,man2023w,wu2024g2ptl,zhu2024unitraj,rose}, leaf, text width=12em]
                        ]
                        [Contrastive
                            [\cite{balsebre2024city,lin2024ptrajm,zhang2024urban}, leaf, text width=12em ]
                        ]
                        [Hybrid
                            [\cite{li2023geolm,dada}, leaf, text width=12em]
                        ]
                    ]
                    [Data Modality
                        [Location
                            [\cite{balsebre2024city,li2022spabert,wu2024g2ptl,zhang2024urban,yan2024urbanclip,hao2024urbanvlp,xiao2024refound}, leaf, text width=12em ]
                        ]
                        [Trajectory/Event
                            [\cite{lin2024trajfm,najjar2023towards,lin2024ptrajm,zhu2024unitraj,chu2023trajgdm,steinberg2023motor}, leaf, text width=12em ]
                        ]
                        [ST Raster
                            [\cite{man2023w,yuan2024urbandit,nguyen2023climax,bi2022pangu,pathak2022fourcastnet,chen2023fengwu,yuan2024unist}, leaf, text width=12em]
                        ]
                        [ST Graph
                            [\cite{shao2022pre,li2024opencity,lam2023learning,prabowo2024traffic,woo2024unified}, leaf, text width=12em]
                        ]
                    ]
                ]
            ]
        \end{forest}
    }    
    \vspace{-1em}
\caption{A method-centric taxonomy. Full version: Fig. \ref{fig:taxonomy_full}.}
\label{fig:taxonomy}
\vspace{-1.5em}
\end{figure}

%


%% file: img/taxonomy_full.tex
\tikzstyle{my-box}=[
    rectangle,
    draw=hidden-draw,
    rounded corners,
    text opacity=1,
    minimum height=1.5em,
    minimum width=5em,
    inner sep=2pt,
    align=center,
    fill opacity=.5,
    line width=0.8pt,
]
\tikzstyle{leaf}=[my-box, minimum height=1.5em,
    fill=hidden-pink!80, text=black, align=left, font=\normalsize,
    inner xsep=2pt,
    inner ysep=4pt,
    line width=0.8pt,
]
\begin{figure*}[!hb]
    \centering
    \resizebox{1 \textwidth}{!}
    {
        \begin{forest}
            forked edges,
            for tree={
                fill=level0!80,
                grow=east,
                reversed=true,
                anchor=base west,
                parent anchor=east,
                child anchor=west,
                base=left,
                font=\large,
                rectangle,
                draw=hidden-draw,
                rounded corners,
                align=left,
                minimum width=4em,
                edge+={darkgray, line width=1pt},
                s sep=3pt,
                inner xsep=2pt,
                inner ysep=3pt,
                line width=0.8pt,
                ver/.style={rotate=90, child anchor=north, parent anchor=south, anchor=center},
            },
            where level=1{text width=2em,font=\normalsize,fill=level1!80,}{},
            where level=2{text width=5em,font=\normalsize,fill=level2!80,}{},
            where level=3{text width=9em,font=\normalsize,fill=level3!60,}{},
            where level=4{text width=10em,font=\normalsize,fill=level4!20,}{},
            where level=5{text width=5em,font=\normalsize,}{},
            [STFM  
                [LLM
                    [Zero-shot \\ Learner
                        [Prompt Engineering
                            [Time-LLM~\cite{jin2023timellm}{,} AuxMobLCast~\cite{xue2022leveraging}{,} Promptcast~\cite{xue2023promptcast}{,} LLMTIME~\cite{gruver2023large}{,} LSTPrompt~\cite{liu2024lstprompt}, leaf, text width=40em ]
                        ]
                        [Agentic Engineering
                            [GeoGPT~\cite{zhang2023geogpt}{,} LLMob~\cite{jiawei2024large}{,} STMA~\cite{lei2025stma}{,} TrafficGPT~\cite{zhang2024trafficgpt}{,} TimeXL~\cite{jiang2025explainable}{,} TimeCAP~\cite{lee2025timecap}, leaf, text width=40em]
                        ]
                    ]
                    [Fine-tuning
                        [All Parameters
                            [W-MAE~\cite{man2023w}{,} ClimaX~\cite{nguyen2023climax}{,} FourcastNet~\cite{pathak2022fourcastnet}{,} GeoLLM~\cite{manvi2023geollm}, leaf, text width=40em ]
                        ]
                        [Partial Parameters
                            [LLM4TS~\cite{chang2023llm4ts}{,} OFA~\cite{zhou2023one}{,} ST-LLM~\cite{liu2024spatial}, leaf, text width=40em ]
                        ]
                        [Add-on Parameters
                            [Time-LLM~\cite{jin2023timellm}{,} LLMLight~\cite{lai2023large}{,} Time-VLM~\cite{zhong2025time}, leaf, text width=40em ]
                        ]
                    ]
                ]
                [PFM
                    [Neural \\ Architecture
                        [Transformer-based
                            [TrajFM~\cite{lin2024trajfm}{,} GeoLM~\cite{li2023geolm}{,} CityFM~\cite{balsebre2024city}{,} OpenCity~\cite{li2024opencity}{,} UniST~\cite{yuan2024unist}{,} ROSE~\cite{rose}, leaf, text width=40em ]
                        ] 
                        [Diffusion-based
                            [TimeDiT~\cite{cao2024timedit}{,} UrbanDiT~\cite{yuan2024urbandit}{,} GPD~\cite{yuan2024spatio}{,} TrajGDM~\cite{chu2023trajgdm}{,} UniMob~\cite{long2024universal}, leaf, text width=40em ]
                        ]
                        [Graph-based
                            [TFM~\cite{wang2023building}{,} GraphCast~\cite{lam2023learning}, leaf, text width=40em ]
                        ]
                        [Others 
                            [TSMamba~\cite{ma2024mamba}{,} Mamba4Cast~\cite{bhethanabhotla2024mamba4cast}{,} Time-SSM~\cite{hu2024time}{,} NeuralGCM~\cite{kochkov2024neural}{,} DADA~\cite{dada}, leaf, text width=40em ]
                        ]
                    ]
                    [Pre-training \\ Scheme
                        [Generative
                            [TrajFM~\cite{lin2024trajfm}{,} TFM~\cite{wang2023building}{,} W-MAE~\cite{man2023w}{,} G2PTL~\cite{wu2024g2ptl}{,} UniTraj~\cite{zhu2024unitraj}{,} ROSE~\cite{rose}, leaf, text width=40em]
                        ]
                        [Contrastive
                            [CityFM~\cite{balsebre2024city}{,} PTrajM~\cite{lin2024ptrajm}{,} MTE~\cite{zhang2024urban}, leaf, text width=40em ]
                        ]
                        [Hybrid
                            [GeoLM~\cite{li2023geolm}{,} DADA~\cite{dada}, leaf, text width=40em]
                        ]
                    ]
                    [Data \\ Modality
                        [Location
                            [CityFM~\cite{balsebre2024city}{,} SpaBERT~\cite{li2022spabert}{,} G2PTL~\cite{wu2024g2ptl}{,} UrbanCLIP~\cite{yan2024urbanclip}{,} UrbanVLP~\cite{hao2024urbanvlp}{,} ReFound~\cite{xiao2024refound}, leaf, text width=40em ]
                        ]
                        [Trajectory/Event
                            [TrajFM~\cite{lin2024trajfm}{,} LTM~\cite{najjar2023towards}{,} PTrajM~\cite{lin2024ptrajm}{,} UniTraj~\cite{zhu2024unitraj}{,}  TrajGDM~\cite{chu2023trajgdm}{,} MOTOR~\cite{steinberg2023motor}, leaf, text width=40em ]
                        ]
                        [ST Raster
                            [W-MAE~\cite{man2023w}{,}~UrbanDiT~\cite{yuan2024urbandit}{,} ClimaX~\cite{nguyen2023climax}{,} Pangu~\cite{bi2022pangu}{,} Fengwu~\cite{chen2023fengwu}{,} UniST~\cite{yuan2024unist}, leaf, text width=40em]
                        ]
                        [ST Graph
                            [STEP~\cite{shao2022pre}{,} OpenCity~\cite{li2024opencity}{,} GraphCast~\cite{lam2023learning}{,} SCPT~\cite{prabowo2024traffic}{,} Moirai~\cite{woo2024unified}, leaf, text width=40em]
                        ]
                    ]
                ]
            ]
        \end{forest}
    }    
    \vspace{-1em}
\caption{Taxonomy from the methodology perspective.}
\label{fig:taxonomy_full}
\vspace{-1.5em}
\end{figure*}
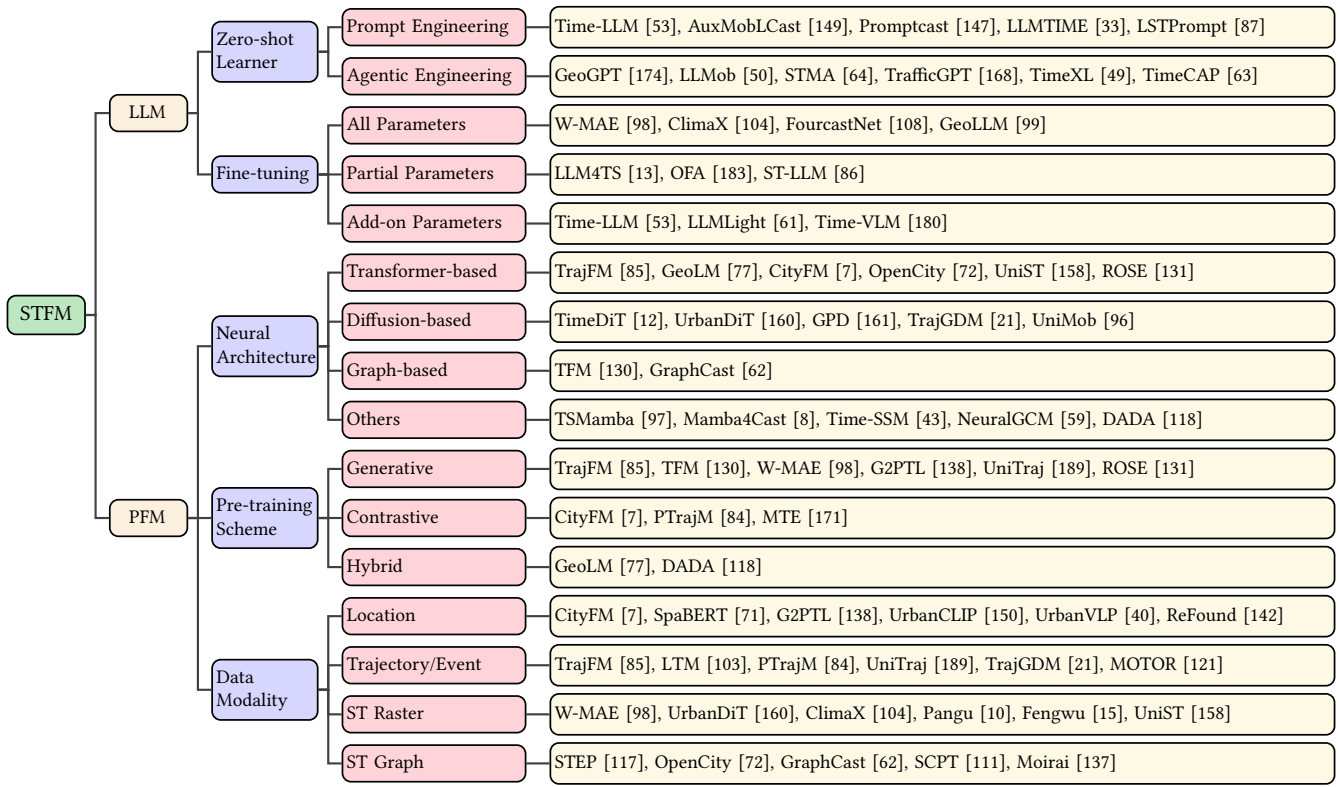

%
